# Symmetry breaking in laser cavities

## Boris A. Malomed

Department of Physical Electronics, School of Electrical Engineering, Tel Aviv University, Tel Aviv 69978, Israel

*A brief introduction to the topic of spontaneous symmetry breaking (SSB) in conservative and dissipative nonlinear systems with an underlying double-well-potential structure is given. The reason is a discussion of a recent observation of the SSB a dual-core nanolaser cavity [5]. The effect is illustrated by means of a simple semi-analytically-tractable model (Fig. 1).*

When discussing the propagation of excitations in a physical system, the shape of the trapping potential determines the system's symmetry. A well-known type is the so-called double-well potential (DWP), which possesses the symmetry due to the use of two adjacent identical wells, see Fig. 1. The DWP is one of the most fundamental settings in quantum mechanics, for which it is commonly known that its ground state (GS) is symmetric, while the first excited state is antisymmetric, both being single (non-degenerate) states [2].

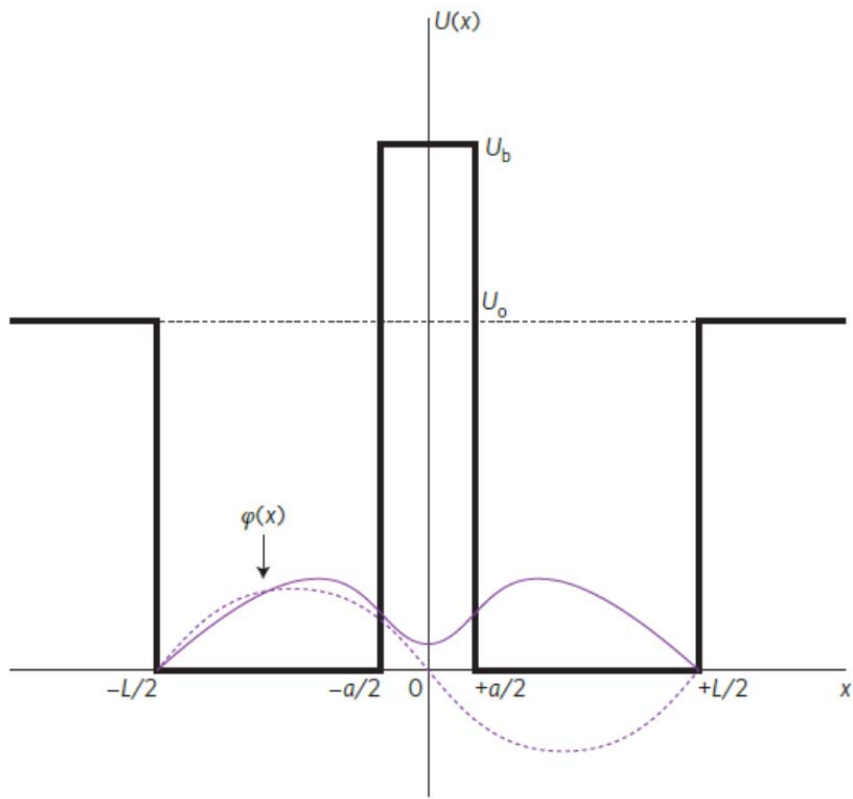

**Figure 1.** The simplest example of the DWP structure in quantum mechanics ($U(x)$) is furnished by a potential box split by a narrow tall barrier in the middle. Even and odd wave functions of the ground and first excited states, $\phi(x)$, are schematically shown by the solid and dashed curves, respectively. In the limit of the infinitely deep potential box, $U_0 \to \infty$, and for the central barrier replaced by the Dirac's delta-function, $\varepsilon \delta(x)$, with $\varepsilon \equiv U_b a$, the GS energy is $\hbar^2 k^2 / (2m)$, where $m$ is the mass of the quantum particle, and $k$ is the smallest root of equation $\tan(kL/2) = -\hbar^2 k / (m\varepsilon)$, while the (larger) energy of the first excited state is $2\pi^2 \hbar^2 / (mL^2)$. In the same case, the ratio of the local minimum of the GS probability at $x = 0$ to its maximum (i.e., the GS modulation depth) is $(\hbar^2 k)^2 / \left[ (m\varepsilon)^2 + (\hbar^2 k)^2 \right]$.

Interestingly, there is a far-reaching analogy between the quantum-mechanical states in a DWP and the propagation of light in two coupled waveguides or cavities, which display a similar symmetry and set of modes. The analogy arises since the behaviours of both systems are governed by similar fundamental equations. The propagation of light is derived from the Maxwell's equations in the paraxial approximation (assuming weak diffraction of a relatively broad beam), which reduces to an equation resembling the linear Schrödinger equation in quantum mechanics. In the presence of optical nonlinearity, the analogy still holds with the nonlinear Kerr effect [1] adding a self-focusing cubic nonlinearity to the propagation equation, making it tantamount to the celebrated nonlinear Schrödinger equation.

In the quantum world, a similar situation occurs when the DWP traps an ultracold rarefied atomic gas in the state of the Bose-Einstein condensate (BEC). In the mean-field approximation, which is extremely accurate for rarefied atomic gases, repulsive or attractive collisions between atoms give rise to a cubic nonlinearity which emulates the optical Kerr effect, with the self-defocusing or focusing sign, respectively. The outcome in this scenario is that the Gross-Pitaevskii equation replaces the linear Schrödinger equation [3] for describing the system.

Importantly, in such systems with the self-focusing nonlinearity the GS symmetry only follows the symmetry of the underlying DWP structure in the weakly nonlinear regime. As the strength of the nonlinearity increases, a fundamental phenomenon called spontaneous symmetry breaking (SSB) occurs [4]. In its simplest form, the SSB implies that an asymmetry develops and the probability to find the quantum particle in one well of the DWP structure, or the intensity of the guided light beam in one core of the dual-core waveguide, is larger than in the other. In the case of the defocusing nonlinearity, the GS symmetry remains unbroken, but the nonlinear term breaks the antisymmetry of the first excited state.

The nonlinear asymmetric GS is double-degenerate: the SSB gives rise to a pair of two mutually symmetric GSs, with the maximum of the atomic density, or the intensity of the guided light beam, spontaneously emerging in either left or right potential well / guiding core.

The same nonlinear system still admits a symmetric state coexisting with the asymmetric ones, but, above the SSB point, the symmetric state no longer represents the GS, being unstable against symmetry-breaking perturbations. Accordingly, in the course of the spontaneous transition from the unstable symmetric state to a stable asymmetric one, the choice between the two mutually degenerate asymmetric states is determined by random perturbations, which "push" the system either to the left or to the right.

Similar to the situation taking place in other areas of physics dominated by nonlinear effects, there is stark imbalance between the number of theoretical and experimental studies on the topic of SSB. While theoretical analyses are numerous and have advanced (see [4] for an overview), there are only a few experimental studies.

An essential step forward in this direction is reported on page xxx of this issue of *Nature Photonics* by Hamel et al. [5]. The researchers demonstrate and analyse SSB in a "photonic molecule" consisting of a dual nanocavity laser embedded into a nonlinear photonic crystal made of a semiconductor material. Such coupled dual cavities are sometimes referred to as a "photonic molecule", as the system's optical modes resemble the electronic states of a diatomic molecule like hydrogen. The two closely-spaced laser cavities are formed by defects in a photonic crystal, with each cavity represented by an area in which three holes in the triangular lattice are missing. The laser emits light at wavelength 1540 nm with gain provided by a layer of quantum dots.

Previous experiments have produced SSB effects in a dual-core guiding structure created in a photorefractive crystal [6], in an atomic BEC loaded into a DWP structure [7], and, recently, in the form of spontaneous breaking of chiral symmetry in metamaterials [8]. SSB in a set of optically coupled semiconductor lasers has also been observed and theoretically modelled [9], but it was caused by a temporal delay in the coupling, which was much larger than the intrinsic oscillation period of each laser.

A peculiarity of the SSB reported in this issue of *Nature Photonics* [5] is that it occurs in a system involving both conservative and dissipative nonlinearities, while most previous theoretical and experimental studies were only dealing with conservative systems. It is worthy to mention that the nonlinearity underlying the operation of the dual-cavity laser reported by Hamel et al [5] is produced not by the Kerr effect (or its varieties, such as the photorefractive nonlinearity [6]), but instead by carrier-density-dependent gain, which is a specific feature of this laser system. This peculiarity is clearly corroborated by the theoretical model of Hamel et al. The model is based on evolution equations for amplitudes of electromagnetic fields in the two coupled cavities, combined with the rate equations for numbers of carriers in them (the model may be simplified, close to the lasing threshold, by effectively eliminating the carrier-population variables, thus reducing the system to a pair of linearly coupled complex Ginzburg-Landau equations (CGLEs)). Simulations of the model demonstrate close agreement with experimental observations.

The basic SSB effect observed and theoretically modelled by Hamel et al [5] is in the form of a pitchfork bifurcation [10], with a spontaneous transition from a symmetric mode, with equal powers in the coupled cavities, to a mode which clearly features a larger power in either of the two cavities. A basic pitchfork-bifurcation diagram, which is common for a broad class of nonlinear dual-cavity systems, including the one realized by Hamel et al, is displayed in Fig. 2(a). This bifurcation, with two mutually symmetric branches of the asymmetric states going forward from the bifurcation (critical) point, belongs to the supercritical (alias forward) type (unlike another generic type, subcritical (alias backward) one, see below, it does not admit coexistence of symmetric and asymmetric states below the critical point). The analysis reported in [5] demonstrates that the observed SSB scenario is quite robust; in particular, it is not altered if random noise is added to the underlying model.

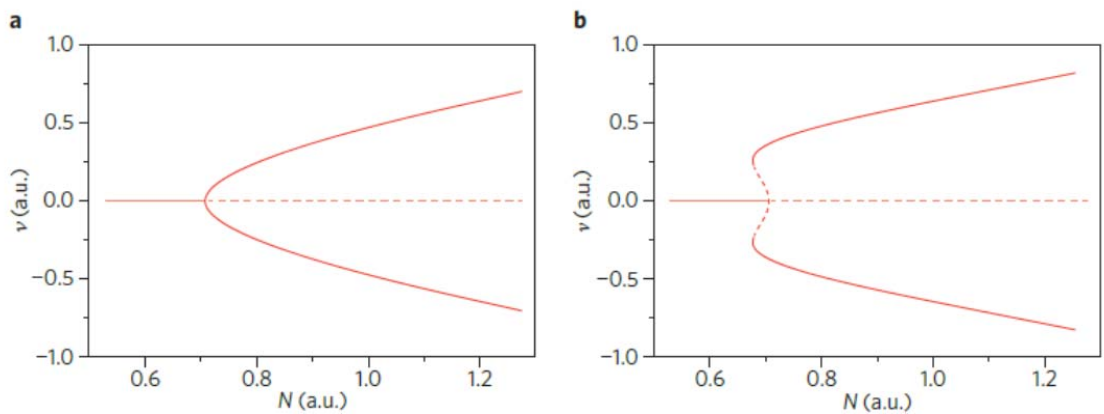

**Figure 2.** (a) The standard form of the supercritical (forward) pitchfork bifurcation in dual-cavity systems (displayed in arbitrary units): with the increase of the nonlinearity strength, represented by the total norm (power) of the trapped physical field, $N$, the imbalance of the norms (powers) between the left and right cores, $\nu = (N_{\text{left}} - N_{\text{right}})/N$, spontaneously jumps from $\nu = 0$, in the stable symmetric state below the respective critical value of $N$, to positive or negative values in stable asymmetric states emerging above the critical point, while the symmetric state becomes unstable (as shown by the dashed line). In particular, for the simplest model presented in Fig. 1, with $U_0 \to \infty$ and norm $N = \int_{-L/2}^{+L/2} \phi^2(x)dx$, its critical value can be easily found for $\varepsilon \gg 2\pi\hbar^2/(mL)$, *viz.*, $N_{\text{cr}} = 8\pi^2\hbar^4/(3mgL^2\varepsilon)$. (b) The generic form of the subcritical (backward) bifurcation. In this case, the dashed backward-going segments of the asymmetric branches are unstable. The same super- and subcritical transitions happen for fixed $N$ if the strength of the linear coupling between the two cores drops below its critical value.

At still higher powers (typically, exceeding the critical value at the SSB point by a factor of $\simeq 1.4$), the asymmetric lasing states observed in [5] demonstrate a secondary instability, accounted for by the Hopf bifurcation [10], which transforms the stationary states into a regime of ultrafast Josephson oscillations between the two cavities, at frequencies of ~150 GHz (Josephson oscillations are closely related to the SSB, leading to an effective dynamical restoration of the broken symmetry [4,7]). These results clearly demonstrate the complexity of the behaviour of nonlinear dissipative systems.

The coexistence of the two robust asymmetric states, which are mirror images of each other, suggests a prospect of using them as elements of a binary-code optical memory. This potential application raises an important question: is it possible to switch one state into another by a control signal? Hamel et al investigate such a possibility in [5] by using a short (~100ps) control laser pulse to selectively illuminate the cavity with the larger lasing intensity in the asymmetric state (while the main pump beam with a central spot 2.2 μm wide is applied symmetrically). The action of the control pulse on carriers in the stronger-excited cavity leads to lasing saturation in it, switching the excitation into the other cavity.

The results reported by Hamel et al provide interesting insights into the behaviour of bifurcations. In particular, it is well known from theoretical studies that the SSB bifurcations feature two generic forms, super- and subcritical ones, alias forward and backward bifurcations (which are tantamount, respectively, to the phase transitions of the second and first kind in statistical physics), as shown in Fig. 2. The subcritical bifurcation features bistability of the symmetric and asymmetric states in a narrow region of the power ( $N$ ) below the critical point. A supercritical bifurcation may be transformed into a subcritical one by a change of the underlying nonlinearity (for example, by a transition from self-focusing to a combined focusing-defocusing nonlinearity).

However, a more relevant option for the expansion of the studies of the SSB phenomenology is to extend the effectively zero-dimensional dual-cavity setting (with each cavity treated as a single quantum dot) to the one-dimensional geometry, in which the two cavities embedded into the photonic crystal are made elongated, in the form of parallel stripes devoid of holes, thus lending the setting a transverse direction, and adding transverse-diffraction terms to the underlying system of coupled CGLEs [11].

In this new geometry, it is predicted that various species of two-component solitons, coupled by super- and subcritical SSB bifurcations, should exist [11]. The creation and experimental observation of such solitons is the next experimental challenge for research in the area. Beyond the framework of photonics, it would be very interesting to see the SSB experimentally realized in a BEC trapped in a dual-core cigar-shaped configuration, as an extension of the single-cigar one where matter-wave solitons have been created [12].

**References**

1. Sauter E. G., *Nonlinear Optics* (John Wiley & Sons: New York, 1996).
2. Landau L. D. and Lifshitz E. M., *Quantum Mechanics* (Nauka Publishers: Moscow, 1974).
3. Dalfovo F., Giorgini S., Pitaevskii L.P., and Stringari S., *Rev. Mod.* Phys. **71**, 463-512 (1999).
4. Malomed, B. A., editor: *Spontaneous Symmetry Breaking, Self-Trapping, and Josephson Oscillations* (Springer-Verlag: Berlin and Heildelberg, 2013).
5. Hamel P., Raineri F., Monnier P., Beaudoin G., Sagnes I., Levenson A., and Yacomotti A. M., *Nature Photonics* **9**, 311-315 (2015).
6. Kevrekidis P. G., Chen Z., Malomed B. A., Frantzeskakis D. J., and Weinstein M. I., *Phys. Lett*. A **340**, 275-280 (2005).
7. Zibold T., Nicklas E., Gross C., and Oberthaler M. K., Phys. Rev. Lett. 105, 204101 (2010).
8. Liu M., Powell D. A., Shadrivov I. V., Lapine M., Kivshar Y. S., *Nature Commun.* **5**, 4441 (2014).
9. Heil T., Fischer I., Elsässer W., Mulet J., and Mirasso C. R., *Phys. Rev. Lett*. **86**, 795-798 (2000).
10. Iooss G. and Joseph D. D., *Elementary Stability and Bifurcation Theory* (Springer-Verlag: New York, 1980).
11. A. Sigler and B. A. Malomed, *Physica D* **212**, 305-316 (2005).
12. Strecker K. E., Partridge B. G., Truscott A. G., and Hulet R. G., *New J. Phys*. **5**, 73.1-73.8 (2003).